\documentclass{ws-p9-75x6-50}

\begin{document}

\title{Bulk Motion Comptonization -- a sure sign of Black Holes}

\author{Sandip K. Chakrabarti and Indranil Chattopadhyay}

\address{S.N. Bose National Centre for Basic Sciences, JD-Block, Salt Lake, Calcutta
700098\\E-mail: chakraba@boson.bose.res.in and indra@boson.bose.res.in}

\maketitle

\abstracts{Bulk motion Comptonization utilizes properties of both matter  and radiation close to the horizon
of a black hole. Computation with these considerations produce hard tails of energy spectral
slope $\sim 1.5-1.7$. These are the most direct evidence of the horizon of a black hole. We argue that even in 
presence of winds and outflows this property is not likely to change as winds are negligible in soft states. }

\noindent Proceedings of 9th Marcel Grossman  Meeting in Rome (Ed. R. Ruffini)

During accretion matter enters the horizon with a velocity of light. This property
has been verified in global solutions of the advective disks\cite{gut}. In the last few Schwarzschild
radii, matter has no time to loss angular momentum since the infall time is
small compared to the viscous timescale to transport angular momentum. 
Thus, angular momentum remains almost constant even in presence of viscosity
and the centrifugal force is very strong compared to gravity at around 10-15 Schwarzschild
radii. Matter slows down, dissipates its radial kinetic energy to thermal energy,
and subsequently proceeds again to be supersonic\cite{chak90}. As matter slows
down, the density rises due to conservation of mass flux and a standing or 
an oscillating shock is formed at this region. Hot, post-shock flow is puffed up
to form a thick accretion disk. Whether it is ion or radiation
pressure supported depends on the net accretion rate of the inflow\cite{mexi93}.
Chakrabarti \& Tititarchuk\cite{ct95} computed the spectra of the
emitted radiation from this centrifugal barrier dominated boundary layer or CENBOL
(See also, Chakrabarti, Titarchuk, Kazanas \& Ebisawa\cite{ctke96}) for various net 
accretion rate which in turn is composed of Keplerian rate (${\dot M}_d$) on the
equatorial plane and sub-Keplerian rate (${\dot M}_h$) away from the equatorial plane.

Chakrabarti \& Titarchuk\cite{ct95} showed that while Keplerian matter in the pre-shock
region is the source of soft photons (soft X-rays), sub-Keplerian
flow supplies hot electrons. Thus, the puffed up post-shock region
is mainly composed of hot electrons which intercepts soft photons from pre-shock region
and Comptonize them to higher energy to produce power-law hard X-ray spectrum. They conclude
that if the soft photons are insufficient to cool down the hot electrons, which is
possible if ${\dot M}_d$ is very small (say ${\dot M}/{\dot M}_{Edd} \sim 0.001 -0.3$)
while ${\dot M}_h \sim 1$, the observed power-law component will be hard and the
black hole will be in a hard state. On the other hand, if ${\dot M}_d$ is high,
say, $\sim 0.4 {\dot M}_{Edd}$ or more, the soft photons intercepted by the post-shock region
totally cools down this region. The emitted spectrum
would be multi-colour black body without any strong power-law hard component. The black hole
would be in a soft state. What they also pointed out is that since this cool matter 
moves almost radially with relativistic speed, they can still Comptonize soft photons
{\it not because of their heat, but because of their bulk motion}.
Only a few photons scattered from these cold but
relativistic electrons escape and produce a 
power-law spectrum with a slope of about $1.5$.

Subsequently, it was shown\cite{chak99} that a significant amount of wind may be
produced from the centrifugal pressure dominated boundary layer. What is more, the ratio
$R_{\dot m}$ of outflow to inflow rates is found to depend on the compression ratio 
$R$ with a peak ($\sim 0.35$) at about $R\sim 2.5$ and going to zero at $R=1$. At large 
$R \sim 4-7$ the ratio falls off to about $0.1$. Though this derivation was obtained assuming 
the outflow to be isothermal till the sonic point, the general nature is found to be
true even for general flow\cite{dc99}. Since in soft states shocks
disappear due to fall of post-shock pressure (see above), $R \sim 1$ and no outflow 
should be produced, while the intermediate states may have very 
high outflow. Hard states should have very low outflows (in absolute terms) even when
sub-Keplerian matter is taken into account.

Meanwhile, Chakrabarti\cite{chak98} computed the spectra of emitted radiation including the
loss of matter from the CENBOL region and found that the spectrum is softened. Later,
Chakrabarti et al.\cite{aa00} claimed that sometimes matter may return from cold
outflows to the CENBOL and this would cause spectral hardening. Indeed, both of these
effects have now been observed (see, Chakrabarti et al.\cite{cetal00}) in the spectra of 
GRS1915+105. However, since in very soft states outflow rate is negligible, there
is no effect of winds on the spectral slopes in this state. So we conclude that the spectral
slope would remain as predicted\cite{ct95}.

Matter entering a black hole has angular momentum.
Rotational motion slows down matter and increases optical depth even at 
lower accretion rates. Thus, there is a distinct effect of angular momentum
on the spectral slopes  in the soft states. For reasonable angular momentum
close to the marginally stable value, it is observed that the slope due to bulk
motion Comptonization should be\cite{ctke96} close to $1.7$ rather than $1.5$ as predicted from
purely radial flow. In several black hole candidates
such high slopes are also observed\cite{etc96}. Good fits of spectra with bulk motion Comptonization 
are presented in Titarchuk et al.\cite{titu98}.


\begin{thebibliography}{99}
\bibitem{gut} S.K. Chakrabarti, {\it Astrophys. J.}, 464, 664 (1996).
\bibitem{chak90} S.K. Chakrabarti, {\it Theory of Transonic Astrophysical Flows} (World Scientific, Singapore, 1990)
\bibitem{mexi93} S.K. Chakrabarti, in {\em  Numerical Simulations in Astrophysics}, eds. J. Franco, S. Lizano,
L. Aguilar \& E. Daltabuit (Cambridge University Press, Cambridge 1993).
\bibitem{ct95} S.K. Chakrabarti and L.G. Titarchuk, {\it Astrophys. J.}, 455, 623 (1995).
\bibitem{ctke96} S.K. Chakrabarti, L.G. Titarchuk, D. Kazanas and K. Ebisawa, {\it Astron. Astrophys. Suppl. Ser.}, 120, 163 (1996).
\bibitem{chak99} S.K. Chakrabarti, 1999, {\it Astron. Astrophys.}, 351, 185 (1999).
\bibitem{dc99} T. Das and  S.K. Chakrabarti, {\it Classical and Quantum Gravity}, 16, 3879 (1999)
\bibitem{chak98} S.K. Chakrabarti, {\it Ind. J. Phys.}, 72B, 565 (1998) astro-ph/9810412.
\bibitem{aa00} S.K. Chakrabarti, S. Manickam, A. Nandi and A.R. Rao, Astron. and Astrophys. (2000) submitted.
\bibitem{cetal00} S.K. Chakrabarti, S. Manickam, A. Nandi and A.R. Rao (2000) this volume.
\bibitem{etc96} K. Ebisawa, L. Titarchuk \& S.K. Chakrabarti, {\it  Publ. Astron. Soc. Jap.}, 48, 59 (1996).
\bibitem{titu98} L.G. Titarchuk et al.  {\em Observational Evidence for Black Holes in the Universe} (Kluwer Academic, Dordrecht 1998)
\end{thebibliography}
\end{document}